\documentclass[namedreferences]{solarphysics}
\usepackage[hyperref,optionalrh,solaromanenum,linksfromyear, optionalrh]{spr-sola-addons}
\usepackage{graphicx}                    
\usepackage{amssymb}                    
\usepackage{color}                       
\usepackage{placeins} 
\usepackage{tabularx}
\usepackage{breakurl}                         

\usepackage{amsmath}
\usepackage{bm}
\usepackage[dvipsnames]{xcolor}
\usepackage{multirow} 
\usepackage{array} 
\makeatletter
\newcommand{\thickhline}{%
    \noalign {\ifnum 0=`}\fi \hrule height 1pt
    \futurelet \reserved@a \@xhline
}
\newcolumntype{"}{@{\hskip\tabcolsep\vrule width 1pt\hskip\tabcolsep}}
\makeatother

\renewcommand{\vec}[1]{ \bm{#1} }
\newcommand{\deriv}[2]{\frac{{\mathrm d} #1}{{\mathrm d} #2}}

\newenvironment{conflicts}[1][Disclosure of Potential Conflicts of Interest]{\footnotesize\paragraph*{#1}}{}

\begin{document}

\begin{article}

\begin{opening}

\title{Analysis of the Helical Kink Stability of Differently Twisted Magnetic Flux Ropes}

%
\author[addressref={aff2},corref,email={marta.florido@hotmail.com}]{\inits{M.}\fnm{M.}~\lnm{Florido-Llinas}\orcid{0000-0001-5834-9836}}

\author[addressref={aff1},email={teresa.nieves@nasa.gov}]{\inits{T.}\fnm{T.}~\lnm{Nieves-Chinchilla}\orcid{0000-0003-0565-4890}}

\author[addressref={aff3},email={mark.linton@nrl.navy.mil}]{\inits{M.G.}\fnm{M. G.}~\lnm{Linton}\orcid{0000-0002-4459-7510}}

\address[id=aff2]{Interdisciplinary Higher Education Center (CFIS), Polytechnic University of Catalonia. Pau Gargallo 14, Barcelona 08028, Spain.}

\address[id=aff1]{Heliophysics Science Division, NASA Goddard Space Flight Center. 8800 Greenbelt Road, Greenbelt MD 20771, USA}

\address[id=aff3]{Naval Research Laboratory. 4555 Overlook Avenue SW, Washington, DC 20375, USA}

%



\runningauthor{Florido-Llinas et al.}
\runningtitle{Helical Kink Stability of Twisted Magnetic Flux Ropes}

\begin{abstract}
Magnetic flux ropes (MFRs) are usually considered to be the magnetic structure that dominates the transport of helicity from the Sun into the heliosphere. They entrain a confined plasma within a helically organized magnetic structure and are able to cause geomagnetic activity. The formation, evolution and twist distribution of MFRs are issues subject to strong debate. Although different twist profiles have been suggested so far, none of them has been thoroughly explored yet. The aim of this work is to present a theoretical study of the conditions under which MFRs with different twist profiles are kink stable and thereby shed some light on the aforementioned aspects. The magnetic field is modeled according to the circular-cylindrical analytical flux rope model in \citeauthor{nieves-chinchilla_circular-cylindrical_2016} (\textit{Astrophys. J.} \textbf{823}, 27, 
\citeyear{nieves-chinchilla_circular-cylindrical_2016}) as well as the Lundquist and Gold-Hoyle models, and the kink stability is analyzed with a numerical method that has been developed based on \citeauthor{linton_helical_1996} (\textit{Astrophys. J.} \textbf{469}, 954, 
\citeyear{linton_helical_1996}). The results are discussed in relation to MFR rotations, magnetic forces, the reversed chirality scenario and the expansion throughout the heliosphere, among others, providing a theoretical background to improve the current understanding of the internal magnetic configuration of coronal mass ejections (CMEs). The data obtained by new missions like Parker Solar Probe or Solar Orbiter will give the opportunity to explore these results and ideas by observing MFRs closer than ever to the Sun.

\end{abstract}

%

\end{opening}

\section{Introduction} \label{s:Introduction}  
 
Coronal mass ejections (CMEs) are large eruptions of magnetized plasma from the solar corona into the heliosphere and one of the main drivers of adverse space weather. They are able to severely impact telecommunications or space systems due to the injection of solar magnetic energy into the magnetosphere, which results from magnetic reconnection processes between the CME and the terrestrial magnetic field. Magnetic flux ropes (MFRs) are fundamental plasma structures that frequently appear in the heliosphere as part of CMEs. They can be defined as collections of magnetic field lines wrapping around an internal main axis in a twisting way, confining magnetized plasma within them.

The in situ signatures of CMEs are known as interplanetary coronal mass ejections (ICMEs). A fraction of the ICMEs detected in the solar wind \citep{gosling_coronal_1990, richardson_fraction_2004} shows the behavior of ideal MFRs: increase in the average magnetic field strength, monotonic rotation of the magnetic field direction through a large angle, low proton temperature, and $\beta_{proton}$ (ratio of plasma to magnetic pressure) significantly lower than 1. These characteristics in the in situ observations were first called ``magnetic clouds" (MCs) by \cite{burlaga_magnetic_1981}. 

Different models have been developed since the early 1980s for the reconstruction of the magnetic field of MFRs in ICMEs only from 1D measurements along the spacecraft observational path. Linear force-free (LFF) configurations with cylindrical geometry \citep{goldstein_field_1983, lundquist_stability_1951, burlaga_magnetic_1988, lepping_magnetic_1990} or toroidal geometry \citep{romashets_force-free_2003} are commonly considered to model MFRs. Nonlinear force-free (NLFF) models are also widely used for this purpose, like the uniformly twisted Gold-Hoyle (GH) model \citep{gold_origin_1960, farrugia_uniform-twist_1999, dasso_new_2006, hu_structures_2014}. Regarding non-force-free (NFF) methods \citep[e.g.][]{mulligan_multispacecraft_2001}, some of them assume a particular current density and then solve Maxwell's equations for the magnetic field with circular or elliptical cross sections \citep{hidalgo_non-force-free_2002, hidalgo_elliptical_2002, nieves-chinchilla_circular-cylindrical_2016, nieves-chinchilla_elliptic-cylindrical_2018}, and there are others like the 2D Grad-Shafranov reconstruction technique \citep{hau_two-dimensional_1999, hu_grad-shafranov_2017}. 

These MFR models differ in the twist they predict (the twist is a quantity that measures how many turns the magnetic field lines make around the axis per unit length). For example, some of them show twist profiles that increase with radius within the cylindrical structure, like the Lundquist model \citep{lundquist_stability_1951} or the most commonly used forms of the circular-cylindrical (CC) analytical model for MCs \citep{nieves-chinchilla_circular-cylindrical_2016}. Other models, like the Gold-Hoyle model (GH: \citealp{gold_origin_1960}), assume a uniform twist.

The kink instability is the canonical instability of twisted magnetic field configurations that makes the flux rope axis become a helix with a pitch similar to the twist of the magnetic field lines. Although the presence of twist is necessary in an MFR to maintain the integrity of the structure \citep{schuessler_magnetic_1979, longcope_evolution_1996}, the instability takes place when the twist becomes larger than a critical value. This stability threshold depends on many factors like the internal magnetic configuration, the external field, the plasma $\beta$ or the aspect ratio, among others \citep{dungey_twisted_1954, hood_kink_1979, mikic_dynamical_1990, linton_helical_1996, bennett_waves_1999, torok_confined_2005}. In the case of a toroidal fusion power reactor, a well-known result is $\Phi_c = 2\pi$, where $\Phi_c$ is the critical total twist angle of a field line around the axis for the occurrence of the kink instability (the \textit{Kruskal-Shafranov limit}, \citealp{shafranov_magnetohydrodynamical_1958, kruskal_hydromagnetic_1958, oz_experimental_2011}). For straight circular-cylindrical geometries, for example \cite{dungey_twisted_1954} and \cite{bennett_waves_1999} showed that the critical twist $\Phi_c$ follows the relation $\Phi_c = 2\frac{L}{R}$ (where $L$ is the axial length and $R$ is the radius of the MFR). Moreover, \cite{hood_critical_1981} analyzed line-tied MFRs (with the footpoints anchored in the photosphere) described by the GH model and found that the line-tying condition had a stabilizing effect, with $\Phi_c = 2.5\pi$. This agrees with the experimental threshold found by \cite{myers_dynamic_2015} in a laboratory set-up resembling solar line-tied MFRs, $\Phi_c \approx 2.5\pi$.

The twist is a relevant property of MFRs since it is related to the amount of magnetic free energy density stored in the structure and its tendency to develop certain instabilities. Studying the twist distribution within ICMEs, and comparing it to the theoretical distributions predicted by different MFR models, could allow us to gain a better insight into which magnetic structures and initiation processes are more feasible. In addition, the analysis of the twist in MFRs could make it possible to predict the range of parameters for which they become kink unstable in the interplanetary medium and possibly start to rotate \citep{vourlidas_first_2011, nieveschinchilla_remote_2012}, just as the kink is already regarded as a promising phenomenon to explain $\delta$-spot rotations during the rise of MFRs through the photosphere \citep{kazachenko_sunspot_2010, vemareddy_sunspot_2016, knizhnik_role_2018}. The analysis of the changes of the kink stability in expanding CMEs \citep{priest_equilibrium_1990, berdichevsky_geometric_2003, berdichevsky_fields_2013} could also provide interesting conclusions about the way CMEs propagate in the interplanetary space.

In the heliosphere, the twist of interplanetary MFRs can be estimated using MFR modeling, along with the Grad-Shafranov reconstruction technique and probes of energetic particles to infer the total field-line length \citep{kutchko_bidirectional_1982, larson_tracing_1997, kahler_solar_2011}. For example, the application of the velocity-modified GH model to 126 MCs in Lepping's list \citep{lepping_summary_2006} showed that all interplanetary MFRs have a twist smaller than $12\pi$ per AU, and their total twist angle $\Phi$ is bounded by $0.2 \frac{L}{R} < \Phi < 2 \frac{L}{R}$ with an average of $\Phi = 0.6 \frac{L}{R}$ \citep{wang_twists_2016}. Some observations support the hypotheses of CMEs being uniformly-twisted structures \citep{farrugia_uniform-twist_1999, hu_magnetic_2015}, or of having a high-twist core enveloped by a less-twisted outer shell \citep{wang_understanding_2018}. However, a recent study based on a superposed epoch analysis of a set of MCs detected by Wind between 1995-2012 shows that the twist distribution is nearly constant in about half their central part (with an average of 11.5 turns per AU), and it increases up to a factor two towards the MC boundaries \citep{lanabere_magnetic_2020}. 
    
The linear force-free (LFF) Lundquist model, which has been widely used to fit a large variety of ICMEs \citep{lepping_summary_2006, lepping_selection_2010, lepping_wind_2018}, shows an increasing twist profile along the radius of the cross section \citep[which may not be infinite in the outer edge as is further discussed in][]{demoulin_re-analysis_2019}. Other NFF models, motivated by the search for less restrictive techniques to reconstruct MFRs \citep{subramanian_self-similar_2014}, are also able to fit great collections of events \citep{nieves-chinchilla_unraveling_2019} while having differing twist distributions.

With regard to CME initiation scenarios, different theories support diverse twist profiles. Some \textit{storage and release} models contemplate the kink or torus instabilities as mechanisms that may play a fundamental role in the initiation of CMEs \citep{linton_relationship_1999, fan_numerical_2004, torok_numerical_2007}. Other hypotheses consider that the eruption is essentially triggered by magnetic reconnection, such as the \textit{flux cancellation} model \citep{moore_filament_1980, van_ballegooijen_formation_1989, moore_onset_2001, amari_coronal_2003} or the \textit{breakout} model \citep{antiochos_model_1999, lynch_observable_2004, sterling_evidence_2004, archontis_eruption_2008}. Part of these models (e.g. flux cancellation) assume the presence of a preexisting MFR \citep{kopp_magnetic_1976, titov_basic_1999} to which magnetic field is subsequently added that could be highly or weakly twisted \citep{longcope_quantitative_2007, aulanier_standard_2012, van_ballegooijen_formation_1989}. This scenario would possibly imply the formation of MFRs with different twist distributions in the core and the outer shell (which will be referred to as \textit{stage-like} twist distributions in this paper). Other models like the breakout explain the formation of a newly developed MFR as a consequence of reconnection processes in conjunction with shearing and twisting motions on the photosphere. Although further research needs to be done on the twist distribution that each initiation possibility would generate, some authors have already started discussing it (e.g. \citealp{wang_understanding_2018}, supports the idea that MFRs have a high-twist core enveloped by a less twisted outer shell, and \citealp{demoulin_re-analysis_2019}, argued that MFRs are created with increasing twist along their cross-sectional radius). The comparison of the CME initiation theories with the heliospheric MFR models in terms of their kink stability and the twist profiles they predict can also provide deeper understanding of the plausibility of different scenarios.

The present work provides a linear kink stability analysis following the procedure in \cite{linton_helical_1996} for MFRs described by the CC model \citep{nieves-chinchilla_circular-cylindrical_2016}, as well as the Lundquist and GH models. In Section \ref{s:Methodology}, the details and behavior of each model and other quantities of interest are described, and the evolution of the CC parameters during expansion is also studied. Section \ref{s:3333} outlines the linear eigenmode analysis and the developed numerical method. Section \ref{s:discussion} shows the results of the analysis, discussing their relation to the occurrence of rotations, magnetic forces, the reversed chirality scenario, the expansion throughout the interplanetary medium, and the shape of the twist profile, among others. A summary is included in Section \ref{s:Summary}.

\section{Theoretical Background and Methodology} \label{s:Methodology}

The methodology followed here is based on the linear stability analysis presented in \cite{linton_helical_1996}. It is applied to a general NFF cylindrical magnetic equilibrium described by the circular-cylindrical analytical model for magnetic clouds (CC) developed in \cite{nieves-chinchilla_circular-cylindrical_2016}, as well as the FF Lundquist and Gold-Hoyle (GH) magnetic models.

\subsection{Magnetic Field Configuration} \label{s:B}

First, the MFR is considered to be an axially symmetric cylinder of cross-sectional radius $R$, with a twisted magnetic field modeled by the CC model \citep{nieves-chinchilla_circular-cylindrical_2016}, which does not impose any condition on the internal magnetic forces. It uses circular-cylindrical coordinates $(r, y, \varphi)$ and basis vectors $\{\vec{e}_r, \vec{e}_y, \vec{e}_\varphi\}$. The CC model assumes that the radial variation of the current density $\vec{j}$ can be expressed as a generic polynomial function,
\begin{equation} \label{eq:current}
    \vec{j} = \sum_{m=0}^\infty \beta_m r^m \vec{e}_y - \sum_{n=1}^\infty \alpha_n r^n \vec{e}_\varphi \ ,
\end{equation}
where $\alpha_n, \beta_m$ are the polynomial coefficients and $r$ is the radial distance from the MFR axis. The magnetostatic Maxwell's equations $\nabla \cdot \vec{B} = 0$ and $\nabla \times \vec{B} = \mu_0 \vec{j}$ are solved in cylindrical coordinates to obtain the general expression of the CC magnetic field, given by Eq. 7 of \cite{nieves-chinchilla_circular-cylindrical_2016}:
\begin{equation} \label{eq:cc_general}
\begin{cases}
    B_r = 0 \\
    \displaystyle B_y = B_y^0 - \mu_0 \sum_{n=1}^\infty \alpha_n \frac{r^{n+1}}{n+1} \\
    \displaystyle B_\varphi = -\mu_0 \sum_{m=0}^\infty \beta_m \frac{r^{m+1}}{m+2}.
\end{cases} \end{equation}
In this work, the polynomial expansion of $\vec{j}$ is truncated to a single term $\alpha_n r^n$ for $j_\varphi$ and $\beta_m r^m$ for $j_y$. This form of the CC model has been chosen because it enables the study of the kink instability including the effects of magnetic forces and different twist distributions that increase along the radius of the MFR. The increasing behavior is supported by observations \citep{lanabere_magnetic_2020} and CME eruption theory \citep{demoulin_re-analysis_2019}. Defining the new variables $\tau = B_y^0/(\mu_0 \alpha_n\frac{1}{n+1}R^{n+1})$, $\tilde{C}_{nm} = -\frac{m+2}{n+1}\frac{\alpha_n}{\beta_m}R^{n-m}$, $\bar{r} = r/R$, the magnetic field components become
\begin{equation} \label{eq:CCAMMC} \begin{cases}
    B_r = 0 \\
    \displaystyle{B_y = B_y^0 \left(1 - \frac{1}{\tau} \bar{r}^{n+1}\right)} \\
    \displaystyle{B_\varphi = \frac{B_y^0}{\tau \tilde{C}_{nm}} \bar{r}^{m+1}}.
\end{cases} \end{equation}
The dependence of the magnetic structure on $\tau$ and $\tilde{C}_{nm}$ will be further discussed in the rest of this section. It is necessary to impose $\lim_{r \to 0} B_\varphi(r) = 0$ and $\lim_{r \to 0} j_\varphi(r) = 0$ to avoid a singularity in the axis, which implies that $n \geq 1$ and $m \geq 0$. In this article, the cases $[n, m] \in \{[1, 0], [2, 1], [3, 2], [1, 1]\}$ will be analyzed. Table \ref{table_eqs} shows the magnetic field expressions, as well as the twist $Q$ and misalignment $\sin\Omega$ that will be explained in Sections \ref{sec:twist} and \ref{sec:FF}. The goal is to find the range of parameters $\tau$ and $\tilde{C}_{nm}$ for which the system is kink unstable for each pair $[n, m]$. The intervals $\tau \in [0.0, 4.0]$ and $\tilde{C}_{nm} \in [0.5, 2.0]$ will be studied since CC fittings often use parameters inside of these ranges.

Secondly, the Lundquist \citep{lundquist_stability_1951, goldstein_field_1983} and GH models \citep{gold_origin_1960} are considered. Their magnetic fields are parametrized as functions of $\alpha$ for the Lundquist model, and $q$ for the GH model (see Table \ref{table_eqs}). The aim will be to find $\alpha$ and $q$, respectively, for which the system is kink stable. 

\begin{table}[h!]
    \centering
    \begin{tabular*}{\textwidth}{clcc}
    
    \textbf{Model} & \multicolumn{1}{c}{$\mathbf{\vec{B}}$} & $\mathbf{Q = \frac{B_\varphi}{rB_y}}$ & $\mathbf{\sin\Omega}$ \\
    \thickhline
    CC $[1, 0]$ & $ \begin{cases}
    {\displaystyle B_y = B_y^0\left(1 - \frac{1}{\tau}\bar{r}^{2}\right)} \\
    {\displaystyle B_\varphi = \frac{B_y^0}{\tau \tilde{C}_{10}}\bar{r} } \end{cases} $ 
    & $\displaystyle \frac{1}{\tilde{C}_{10}R ( \tau - \bar{r}^2 )}$
    & $\displaystyle \frac{\bar{r}\left(\tau - \bar{r}^2 - \frac{1}{\tilde{C}_{10}^2}\right)}{\left[(\tau - \bar{r}^2)^2 + \frac{\bar{r}^2}{\tilde{C}_{10}^2}\right]^\frac{1}{2} \left[\bar{r}^2 + \frac{1}{\tilde{C}_{10}^2}\right]^\frac{1}{2}} $ \\
    \hline
    CC $[2, 1]$ & $ \begin{cases}
    {\displaystyle B_y = B_y^0\left(1 - \frac{1}{\tau}\bar{r}^{3}\right)} \\
    {\displaystyle B_\varphi = \frac{B_y^0}{\tau \tilde{C}_{21}}\bar{r}^2 } \end{cases} $ 
    & $\displaystyle \frac{\bar{r}}{\tilde{C}_{21}R ( \tau - \bar{r}^3 )}$  
    & $\displaystyle \frac{\bar{r}\left(\tau - \bar{r}^3 - \frac{\bar{r}}{\tilde{C}_{21}^2}\right)}{\left[(\tau - \bar{r}^3)^2 + \frac{\bar{r}^4}{\tilde{C}_{21}^2}\right]^\frac{1}{2} \left[\bar{r}^2 + \frac{1}{\tilde{C}_{21}^2}\right]^\frac{1}{2}} $ \\
    \hline
    CC $[3, 2]$ & $ \begin{cases}
    {\displaystyle B_y = B_y^0\left(1 - \frac{1}{\tau}\bar{r}^{4}\right)} \\
    {\displaystyle B_\varphi = \frac{B_y^0}{\tau \tilde{C}_{32}}\bar{r}^3 } \end{cases} $ 
    & $\displaystyle \frac{\bar{r}^2}{\tilde{C}_{32}R ( \tau - \bar{r}^4 )}$ 
    & $\displaystyle \frac{\bar{r}\left(\tau - \bar{r}^4 - \frac{\bar{r}^2}{\tilde{C}_{32}^2}\right)}{\left[(\tau - \bar{r}^4)^2 + \frac{\bar{r}^6}{\tilde{C}_{32}^2}\right]^\frac{1}{2} \left[\bar{r}^2 + \frac{1}{\tilde{C}_{32}^2}\right]^\frac{1}{2}} $ \\
    \hline
    CC $[1, 1]$ & $ \begin{cases}
    {\displaystyle B_y = B_y^0\left(1 - \frac{1}{\tau}\bar{r}^{2}\right)} \\
    {\displaystyle B_\varphi = \frac{B_y^0}{\tau \tilde{C}_{11}}\bar{r}^2 } \end{cases} $ 
    & $\displaystyle \frac{\bar{r}}{\tilde{C}_{11}R ( \tau - \bar{r}^2 )}$  
    & $\displaystyle \frac{\tau - \bar{r}^2 \left( 1 + \frac{3}{2\tilde{C}_{11}^2} \right)}{\left[(\tau - \bar{r}^2)^2 + \frac{\bar{r}^4}{\tilde{C}_{11}^2}\right]^\frac{1}{2} \left[1 + \frac{1}{\tilde{C}_{11}^2}\right]^\frac{1}{2}} $ \\
    \hline
    Lund. & $ \begin{cases}
    {\displaystyle B_y = B_y^0 J_0(\alpha \bar{r})} \\
    {\displaystyle B_\varphi = B_y^0 J_1(\alpha \bar{r})} \end{cases} $ 
    & $\displaystyle \frac{J_1(\alpha\bar{r})}{R\bar{r}J_0(\alpha\bar{r})}$  
    & $0$ \\
    \hline
    GH & $ \begin{cases}
    {\displaystyle B_y = \frac{B_y^0}{1 + q^2 \bar{r}^2}} \\
    {\displaystyle B_\varphi = \frac{q\bar{r}B_y^0}{1+q^2\bar{r}^2}} \end{cases} $ 
    & $\displaystyle \frac{q}{R}$ & $0$ \\
    
    \end{tabular*}
    \caption{Magnetic field, twist and misalignment equations (see Eq. \ref{eq:misalignment} for the definition of the force-free measure given by the misalignment $\sin\Omega$).}
    \label{table_eqs}
\end{table}

\begin{figure}[h!]
    \centering
    \includegraphics[width=0.66\textwidth]{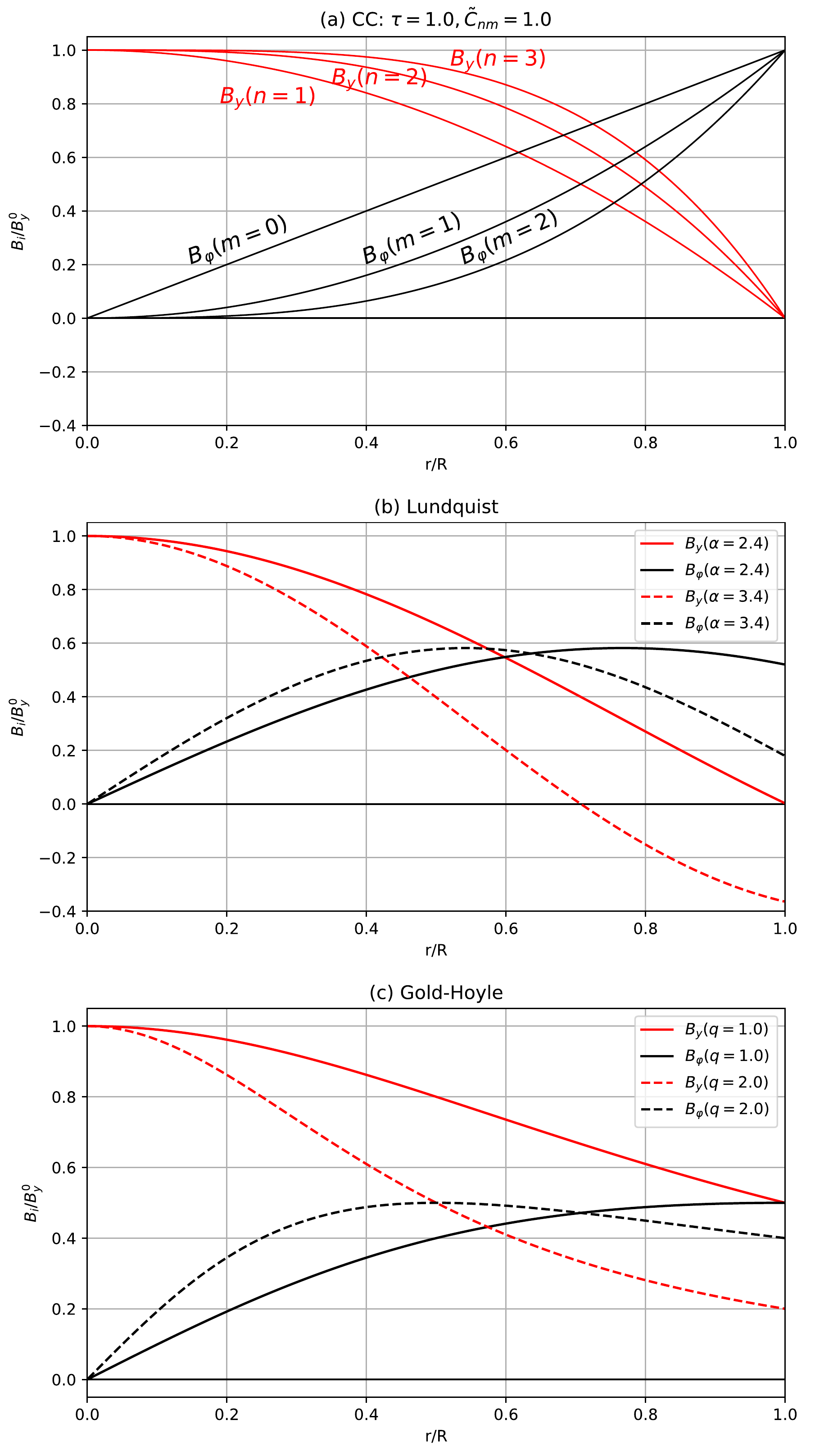}
    \caption{Axial and poloidal magnetic field components of (a) the CC model for $[n, m] = [1, 0], [2, 1], [3, 2], [1, 1]$ with parameters $\tau = 1.0, \tilde{C}_{nm} = 1.0$; (b) the Lundquist model for $\alpha = 2.0, 3.0$; (c) the GH model for $q = 1.0, 2.0$.}
    \label{fig_B}
\end{figure}

Figure \ref{fig_B}a shows the normalized CC magnetic field components $B_y$ (in red) and $B_\varphi$ (in black) along the normalized radius of the MFR for $\tau = 1.0$ and $\tilde{C}_{nm} = 1.0$. The axial magnetic field $B_y$ is set to $B_y^0$ in the core and decreases towards the boundary of the MC by an amount that depends on the parameter $\tau$: it vanishes there if $\tau = 1$, becoming negative when $\tau < 1$ or closer to $B_y^0$ if $\tau > 1$. The poloidal component $B_\varphi$ is zero in the axis and grows towards the outer edge with a rate of change that is inversely proportional to $\tilde{C}_{nm}$ and $\tau$.

Figure \ref{fig_B}b and \ref{fig_B}c show the Lundquist and GH magnetic fields for $\alpha = 2.4, 3.4$ and $q = 1.0, 2.0$, respectively. In fact, varying $\alpha$ in the Lundquist model only squeezes or stretches the horizontal axis, and so it indicates where the cutoff point of the Bessel functions is taken as the MFR outer radius. The GH model behaves similarly. For both models, this entails a bigger difference between $B_y$ in the core and the boundary when $\alpha$ or $q$ are increased. For the Lundquist case, $B_y$ becomes negative when $\alpha > 2.4$. In the GH model, $B_y$ stays always positive. 

\subsubsection{Twist or Helical Pitch} \label{sec:twist}

The number of turns of the MFR in an axial length $L$ is given by $N_L(r) = | Q L / 2\pi |$. $Q$ will be referred to as the \textit{twist} or \textit{helical pitch} of the magnetic structure and it can be interpreted as a wavenumber measuring the angle covered by magnetic field lines per unit length. It is given by
\begin{equation} \label{eq:twist}
    Q = \frac{B_\varphi}{r B_y}.
\end{equation}

\begin{figure}[h!]
    \centering
    \includegraphics[width=1.0\textwidth]{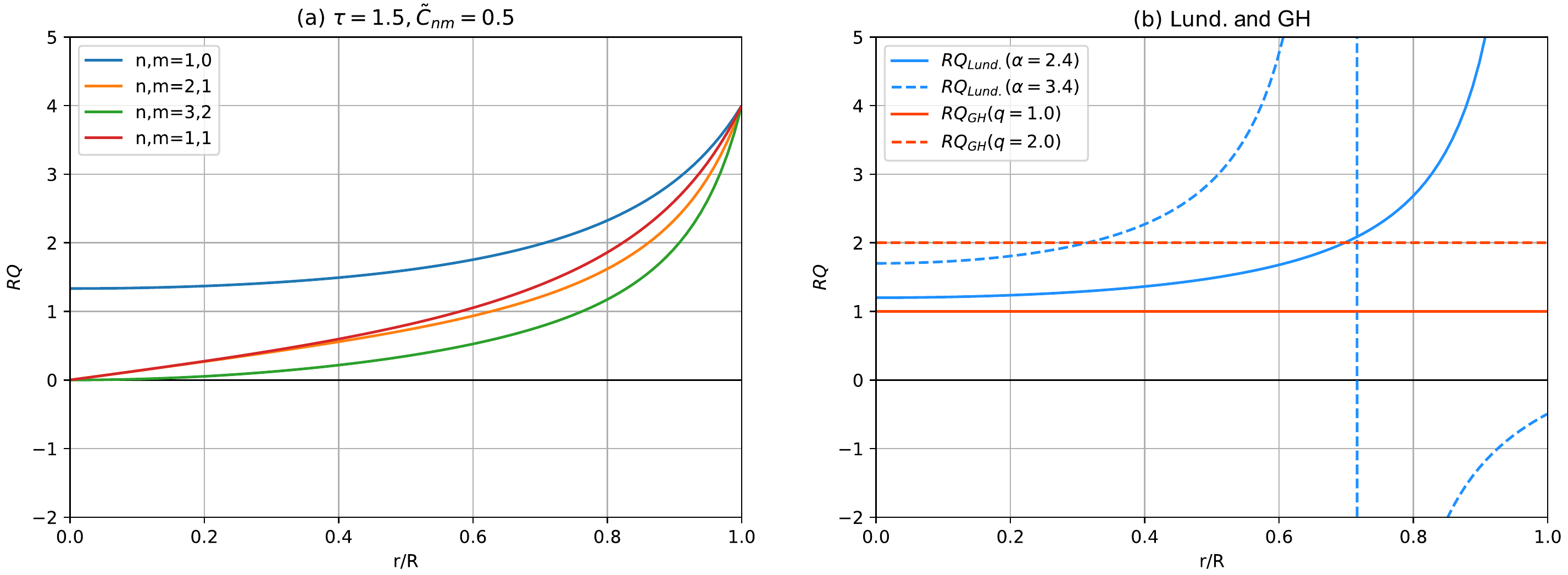}
    \caption{Measurement of $RQ$ along the flux rope radius for (a) CC $[n, m] = [1, 0], [2, 1], [3, 2], [1, 1]$ with parameters $\tau = 1.5, \tilde{C}_{nm} = 0.5$; (b) the Lundquist model (in blue) for $\alpha = 2.4$ (continuous) and $3.4$ (dashed), and the GH model (in orange) for $q = 1.0$ (continuous) and $q = 2.0$ (dashed).}
    \label{fig_Q}
\end{figure}

Figure \ref{fig_Q}a shows the behavior of the product $RQ$ along the normalized radius for parameters $\tau = 1.5$ and $\tilde{C}_{nm} = 0.5$. For the CC cases of the form $[n, 0]$, $Q$ adopts the finite value of $1/(\tilde{C}_{n0}R\tau)$ in the core, while it vanishes for the rest of the other $[n, m]$ forms. Then, it increases towards the boundary reaching a value that depends on $\tau$: if $\tau = 1$ it goes to infinity in the edge; if $\tau < 1$ the twist becomes infinite at an internal point of the MFR ($\bar{r} = \tau$) and then reverses its sign causing the chirality to change; and if $\tau > 1$, the twist goes to $1/(\tilde{C}_{nm}R(\tau - 1))$ at the edge. The twist $Q$ is inversely proportional to $\tilde{C}_{nm}$ and $R$, and the $Q$ profile decreases with increasing $\tau$ and adopts a more uniform shape, so that it is equal to $0$ in the limit $\tau \to \infty$. Moreover, for cases of the form $[k+1, k]$, larger $k$ implies a smaller growth rate of the twist around the core, thus adopting more of a stage-like distribution, or in other words, an MFR with a twist distribution around the core that is different from the one in its outer shell (in this case, it would be almost uniform in the core, and abruptly increase close to the boundary). 

Figure \ref{fig_Q}b displays in blue the quantity $RQ$ in the Lundquist model for $\alpha = 2.4, 3.4$. It always has an increasing profile, growing towards a finite value in the boundary if $\alpha < 2.4$ and to infinity if $\alpha = 2.4$. When $\alpha > 2.4$ there is a change in the chirality of the MFR that occurs at $\bar{r} = 2.4/\alpha$. Figure \ref{fig_Q}b also shows in orange the uniform twist $RQ$ of GH model for $q = 1.0, 2.0$. 

\subsubsection{Misalignment Between $\vec{j}$ and $\vec{B}$} \label{sec:FF}

An MFR is said to be \textit{force-free} (FF) if the magnetic field is completely aligned with the current density ($\vec{j}\times\vec{B} = 0$), such that $\nabla \times \vec{B} = \alpha \vec{B}$. Lundquist and GH models are FF. A measure of the \textit{force-freeness} for the CC model (Eq. \ref{eq:CCAMMC}) is given by the misalignment between $\vec{j}$ and $\vec{B}$,
\begin{align} 
    \sin\Omega &= \frac{(\vec{j} \times \vec{B})\vert_r}{|\vec{j}||\vec{B}|}  \nonumber \nonumber \\ 
    &= \frac{ \bar{r}^n (\tau - \bar{r}^{n+1}) - \frac{m+2}{n+1}\frac{\bar{r}^{2m+1}}{\tilde{C}_{nm}^2}}{ \left[ (\tau - \bar{r}^{n+1})^2 + \frac{1}{\tilde{C}_{nm}^2} \bar{r}^{2(m+1)} \right]^{\frac{1}{2}} \left[ \bar{r}^{2n} + \frac{1}{\tilde{C}_{nm}^2} \bar{r}^{2m} \right]^{\frac{1}{2}} }. \label{eq:misalignment}
\end{align}
The configuration is FF in the core for the cases with $n > m$. An MFR modeled by Eq. \ref{eq:CCAMMC} is considered to be \citep{nieves-chinchilla_circular-cylindrical_2016}:
\begin{itemize}
    \item Non-FF (NFF), Lorentz force pointing outwards, if $\sin\Omega > 0.3$ ($\Omega > 18^\circ$).
    \item FF, if $|\!\sin\Omega| < 0.3$ ($|\Omega| < 18^\circ$).
    \item NFF, Lorentz force pointing inwards, if $\sin\Omega < -0.3$ ($\Omega < -18^\circ$).
\end{itemize} 

\begin{figure}[h!]
   \centerline{\includegraphics[width=1.0\textwidth]{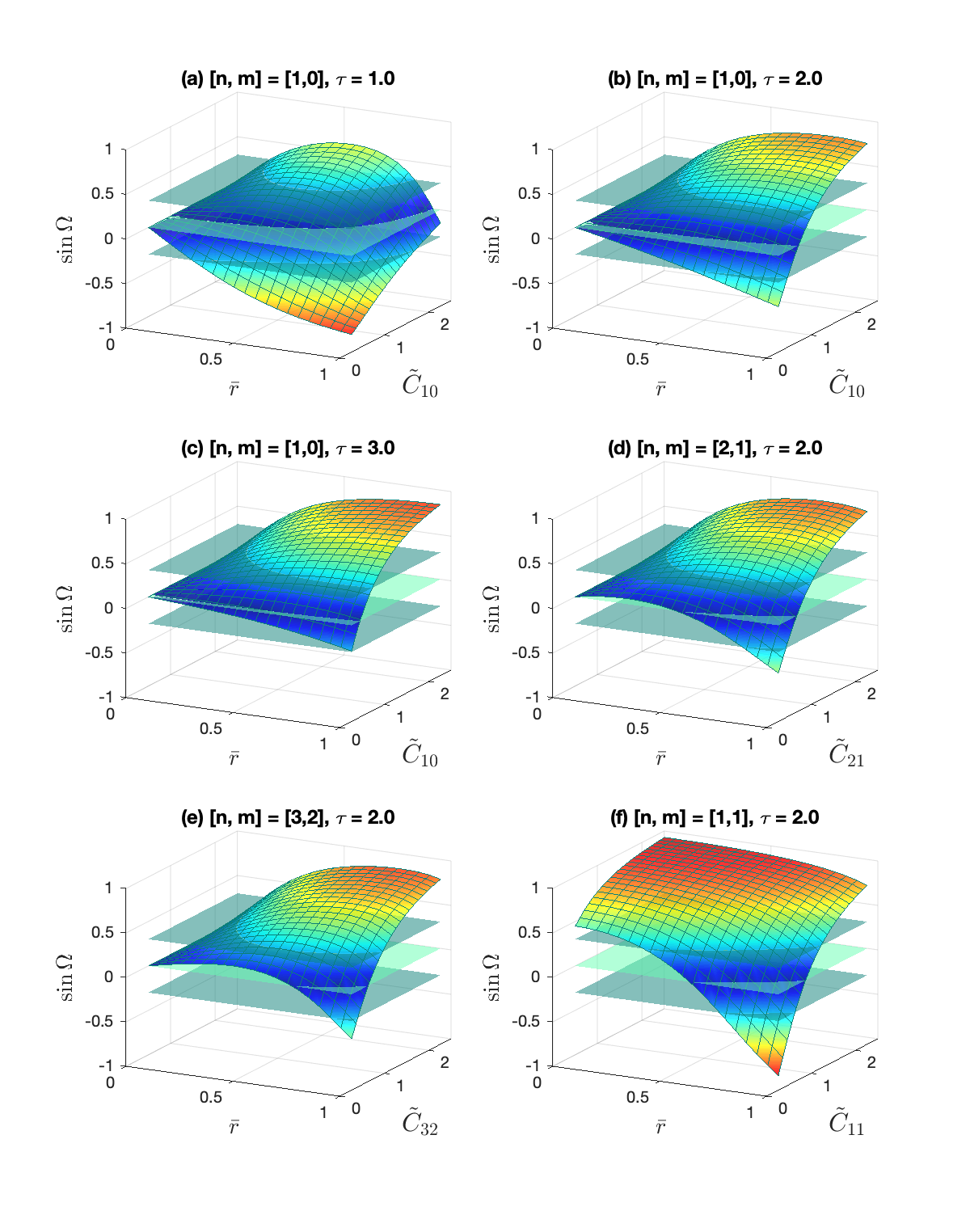}}
   \caption{Plots of the misalignment $\sin\Omega$ in function of the normalized radius and the parameter $\tilde{C}_{nm}$. (a), (b), and (c) show the variation of $\sin\Omega$ with increasing $\tau$ for $[1, 0]$. (b), (d), and (e) show how $\sin\Omega$ changes when $k$ in $[k+1, k]$ is increased for constant $\tau = 2.0$. The outward-NFF behavior of the case [1, 1] in the core is displayed in (f).}
   \label{fig_FF}
\end{figure}

Figure \ref{fig_FF} displays how the misalignment $\sin\Omega$ between the magnetic field and the current density vector varies along the radius and with the parameter $\tilde{C}_{nm}$, for different pairs $n, m$, and $\tau$. Three planes corresponding to $\sin\Omega = -0.3, 0, 0.3$ are shown in each of them. 

In general, it can be observed that smaller $\tilde{C}_{nm}$ makes the structure inward-NFF in the vicinity of the boundary, while bigger $\tilde{C}_{nm}$ makes it be outward-NFF in the middle and outer sections. The top three panels in Figure \ref{fig_FF} show how the misalignment changes in the case $[1,0]$ as $\tau$ is increased: larger $\tau$ makes the structure be more FF (for small $\tilde{C}_{nm}$) or outward-NFF towards the edge (for big $\tilde{C}_{nm}$), so that the inward-NFF behavior disappears. This is valid for the rest of $[n, m]$. Moreover, the case $[1, 1]$ (see Figure \ref{fig_FF}f) has been included in the analysis because it is outward-NFF in the core, while the ones of the form $[k+1, k]$ are FF in the axis and become slightly more outward-NFF in the mid and outer sections as $k$ is increased (see Figure \ref{fig_FF}b, \ref{fig_FF}d, \ref{fig_FF}e for fixed $\tau = 2.0$).

\subsubsection{Expansion} \label{s:expansion}

In the ICME journey throughout the heliosphere, the relative magnetic helicity $H_r$ and magnetic fluxes $\phi_y, \phi_\varphi$ are conservative quantities if there is no erosion or reconnection with the ambient solar wind. The expressions for these physical quantities for the CC model are given in Eq. 5, 6, and 11 of \cite{nieves-chinchilla_modeling_2018} and reproduced here:
\begin{align*}
    \phi_y &= \pi R^2 B_y^0 \left[ 1 - \frac{2}{(n+3)\tau} \right] \\
    \phi_\varphi &= LR \frac{B_y^0}{(m+2)\tau\tilde{C}_{nm}} \\
    \frac{H_r}{L} &= \frac{4\pi R^3(B_y^0)^2}{\tau\tilde{C}_{nm}} \left[ \frac{1}{2(m+4)} - \frac{1}{(n+3)(m+n+5)\tau} \right],
\end{align*}
where the relative helicity is expressed per unit length. It is assumed that the MFR expands from a radius $R$ to $R'$, and from an axial length $L$ to $L'$. The CC parameters when the MFR has radius R ($\tau, \tilde{C}_{nm}$, and $B_y^0$) are known. The aim is to find the parameters $\tau', \tilde{C}'_{nm}$, and ${B_y^0}'$ of the MFR when it expands to a radius $R'$ and axial length $L'$, in terms of $\tau, \tilde{C}_{nm}$, and $B_y^0$. This is achieved by making equal the magnetic fluxes and the relative helicity at the two evolutionary stages: $\phi_y = \phi_y', \phi_\varphi = \phi_\varphi'$, and $H_r = H_r'$. Isolating $1/\tau'$ in $\phi_y = \phi'_y$, one obtains
\begin{equation} \label{eq:1/tau}
    \frac{1}{\tau'} = \frac{n+3}{2} + \frac{B_y^0}{{B_y^0}'}\left(\frac{R}{R'}\right)^2\left[\frac{1}{\tau} - \frac{n+3}{2}\right],
\end{equation}
and $\phi_\varphi = \phi_\varphi'$ gives
\begin{equation} \label{eq:Cnm/Cnm'}
    \frac{\tilde{C}'_{nm}}{\tilde{C}_{nm}} = \frac{L'}{L}\frac{R'}{R}\frac{{B_y^0}'}{B_y^0} \frac{\tau}{\tau'}
\end{equation}
Substituting Eq. \ref{eq:1/tau} and \ref{eq:Cnm/Cnm'} in $H_r = H'_r$, the final results are obtained:
\begin{equation} \begin{cases} \label{eq:expansion}
    \displaystyle {B_y^0}' = \left( \frac{R}{R'} \right)^2 B_y^0 \\
    \tau' = \tau \\
    \displaystyle \tilde{C}_{nm}' = \frac{L'}{L} \frac{R}{R'} \tilde{C}_{nm} \ .
\end{cases} \end{equation}
This means that upon expansion, $\tau$ remains constant (so the ratio of $B_y$ in the core to its value in the boundary does not change). $B_y^0$ decreases in a way inversely proportional to $(R')^2$, and $\tilde{C}_{nm}$ can increase or decrease depending on the relation between $\frac{L'}{L}$ and $\frac{R'}{R}$. In terms of the magnetic field components and twist,
\begin{equation*} \begin{cases}
    \displaystyle B_y' = \left( \frac{R}{R'} \right)^2 B_y \\
    \displaystyle B_\varphi' = \frac{R}{R'}\frac{L}{L'}B_\varphi \\
    \displaystyle Q' = \frac{L}{L'}Q \ ,
\end{cases} \end{equation*} 
which implies that the twist will decrease if the MFR axial length increases.


\section{Analysis} \label{s:3333}

\subsection{Linear Stability Analysis}

The kink instability is studied using the energy principle method \citep{bernstein_energy_1958} as developed in \cite{linton_helical_1996}, which evaluates the linear stability of an MHD equilibrium that undergoes a small displacement perturbation $\vec{\xi}$. An outline of this method will be given in the present section. It depends upon a variational formulation of the equations of motion of the plasma, and aims at discovering whether there is any perturbation that decreases the potential energy from its equilibrium value, thus making the system unstable. The cylindrically symmetric magnetic equilibrium (with radius $R$) under study is considered to be surrounded by field-free plasma, and confined by the higher pressure of this external plasma (for further details on the energy principle, see e.g. \citealp{bernstein_energy_1958, bateman_mhd_1978, lifschitz_magnetohydrodynamics_1989}).
    
The starting point is to linearize the system of ideal MHD equations and boundary conditions by considering an arbitrarily small perturbation $\bm{\xi}(\vec{r}, t)$ from a state of stationary equilibrium. They are then combined into a single second order partial differential equation for the displacement vector $\bm{\xi}(\vec{r}, t)$ that is expressed as \begin{equation} \rho_0 \ddot{\bm{\xi}} = \vec{F}(\bm{\xi}), \label{eq:motion} \end{equation} where $\rho_0$ is the equilibrium density of the plasma and $\vec{F}$ is the self-adjoint operator that represents the force per unit volume in the plasma. The initial conditions are $\bm{\xi}(\vec{r}, 0) = \vec{0}$ and $\dot{\bm{\xi}}(\vec{r}, 0) = \bm{\xi}_0$. The potential energy of the system is $$ W = -\frac{1}{2} \int d^3x  \, \bm{\xi}^* \cdot \vec{F}(\bm{\xi}), $$ where $\bm{\xi}^*$ is the complex conjugate of $\bm{\xi}$. The energy principle as derived by \cite{bernstein_energy_1958} states that the system is stable if and only if $W \geq 0$ for all possible perturbations $\bm{\xi}$. In the method of \cite{linton_helical_1996}, $W$ is extremized while the integral $$ K = \frac{1}{8\pi}\int d^3 x \, |\bm{\xi}|^2$$ is kept constant. This constrained variation can be done by finding the extrema of a generalized energy $U = W + \lambda K$, where $\lambda$ is a Lagrange multiplier and $K$ is fixed. This is equivalent to finding the eigenvalues of $\vec{F}$ assuming a time dependence of $\bm{\xi} \propto e^{i\omega t}$, since Eq. \ref{eq:motion} does not depend explicitly on time (see e.g. \citealp{chiappinelli_nonlinear_2019}). Then, $\lambda$ is related to $\omega$ by $$ \lambda = -4\pi \rho_0 \omega^2, $$ and $\omega^2 \in \mathbb{R}$ because $\vec{F}$ is a self-adjoint operator. The system will develop a kink instability of growth rate $|\omega|$ if and only if $\omega^2 < 0$ (positive $\lambda$) for every eigenvalue of $\vec{F}$.
    
    
    To study the kink mode, the displacement perturbation is assumed to have helical symmetry with wavenumber $k$ and arbitrary radial structure, $\vec{\xi}(\vec{r}) = [\xi_r(r), \xi_y(r), \xi_\varphi(r)]e^{i(k y + \varphi)}$. The perturbation that minimizes the generalized energy of the MFR can be obtained from the radial component $\xi_r$ given by the Euler-Lagrange equation
    \begin{equation}\label{eq_euler}
        \deriv{}{r} \left( f \deriv{\xi_r}{r} \right) - g\xi_r = 0,
    \end{equation}
    where $f$ and $g$ are defined as
    $$ f = \frac{r^3(\lambda + \left( kB_y + \frac{B_\varphi}{r} \right)^2}{1 + k^2 r^2}, $$
    \begin{align*} g = & \frac{k^2 r}{1 + k^2 r^2} \left[ r^2 \left\{ \lambda + \left( kB_y + \frac{B_\varphi}{r} \right)^2 \right\} - r\deriv{|\vec{B}(r)|^2}{r} \vphantom{\frac{2\left( kB_y + \frac{B_\varphi}{r} \right)^2}{\lambda + \left( kB_y + \frac{B_\varphi}{r} \right)^2}}\right. \\ 
     & \left. - 2B_\varphi^2  \left\{ \frac{2\left( kB_y + \frac{B_\varphi}{r} \right)^2}{\lambda + \left( kB_y + \frac{B_\varphi}{r} \right)^2} - 1 \right\} + \frac{2}{1 + k^2 r^2} (r^2\lambda + k^2 r^2 B_y^2 - B_\varphi^2) \right]. \end{align*}
    Regularity at the origin is ensured by the boundary conditions $\dot{\xi}_r(0) = 0$ and $\xi_r(0) = \xi_0$ ($\xi_0$ can be set to 1 without loss of generality). Other stability analyses of tokamaks or coronal loops have assumed the confinement of the tube by a conducting wall ($\xi_r(R) = 0$, see e.g. \citealp{voslamber_stability_1962}) or the presence of an external vacuum field $B(r > R) \neq 0$ (e.g. \citealp{kruskal_hydromagnetic_1958, hood_critical_1981}). In contrast, in this work, it is considered that the MFR has a free boundary and no external magnetic field \citep{linton_helical_1996}, since it is an approximation to the most common scenario in the heliosphere. Future studies should regard some of the aforementioned different boundary conditions to account for the occurrence of interaction phenomena. An additional condition, $D(\lambda; R, k) = 0$, is therefore obtained when imposing the continuity of the total pressure across the boundary and the Euler-Lagrange equation at the outer edge of the tube, even if there is a discontinuity in the magnetic field. The function $D(\lambda; R, k)$ can be regarded as a dispersion function for the eigenvalue $\lambda$, and is given by the expression
    \begin{align} D(\lambda; R, k) = \left[ k^2 |\vec{B}(R)|^2 + \lambda + \lambda \frac{(1 + k^2 R^2) K_1(|k|R)}{|k|RK_0(|k|R) + K_1(|k|R)} \right] &\xi_r(R) \nonumber \\ + \left\{ R\lambda + R\left(kB_y + \frac{B_\varphi}{r} \right)^2 \right\} &\dot{\xi}_r(R), \label{eq:disp}  \end{align}
    where $K_0$ and $K_1$ are modified Bessel functions. A circular-cylindrical MFR with given $R$ and $\vec{B}(r)$ is said to be kink stable if it is stable to perturbations of any wavenumber $k$, so a necessary and sufficient condition for the kink stability is that the largest $\lambda$ for which the dispersion relation in Eq. \ref{eq:disp} holds, keeps on being negative for all $k$.

\subsection{Numerical Analysis} \label{s:Numerical analysis}

Given a particular pair $[n, m]$ defining the magnetic equilibrium $\vec{B}_0(r)$ for the CC model, the purpose of the numerical procedure that has been developed is to find the value of $\tau$ for each $\tilde{C}_{nm}$ above which the system becomes kink stable to perturbations of any wavenumber $k$, called $\tau_{crit}$. It uses Brent's method \citep{brent_algorithms_2013} to find the zeros of the dispersion relation with appropriately chosen bracketing intervals. Each evaluation of $D(\lambda; R, k)$ requires solving an Euler-Lagrange equation, and this has been implemented with the \texttt{odeint} Python routine (source code from \citealp{oliphant_scipyintegrateodepackpy_2019}), which applies the Livermore Solver for Ordinary Differential Equations (LSODA). The problem is solved for the dimensionless quantities $\tilde{\lambda} = \lambda R^2/(B_y^0)^2$, $\tilde{k} = kR$, $\bar{r} = r/R$, and $\tilde{\vec{B}} = \vec{B}/B_y^0$.

For fixed $\tau$ and $\tilde{C}_{nm}$, the program needs a first guess of $\tilde{k}_{min}$, $\tilde{k}_{max}$ (the minimum and maximum $\tilde{k}$ for which the largest $\tilde{\lambda}$ that solves $D(\tilde{\lambda}; 1, \tilde{k}) = 0$ is positive), and also the largest zero of $D(\tilde{\lambda}; 1, \tilde{k})$ only for one arbitrary value of $\tilde{k}$. The output consists of the precise values of $\tilde{k}_{min}$, $\tilde{k}_{max}$, $\tilde{\lambda}_{ext}$ (where $\tilde{\lambda}_{ext} = \max\{ \tilde{\lambda} > 0 \mid \exists \, \tilde{k} \text{ s.t. } D(\tilde{\lambda}; 1, \tilde{k}) = 0\}$) and $\tilde{k}_{ext}$ (wavenumber for which the largest zero of $D(\tilde{\lambda}; 1, \tilde{k}_{ext})$ is $\tilde{\lambda}_{ext}$). For a given $\tilde{C}_{nm}$, these parameters are found for three different $\tau$ with $\tilde{\lambda}_{ext}$ of the order of $10^{-10} - 10^{-8}$ by providing the aforementioned first guesses by graphical inspection. The program then outputs $\tilde{k}_{min}, \tilde{k}_{max}, \tilde{k}_{ext}, \tilde{\lambda}_{ext}$ for a desired number $n$ of points in the range of the given values of $\tau$. Next, a decreasing exponential function is fitted to $\tilde{\lambda}_{ext}$ as a function of $\tau$. The criterion chosen to define the $\tau_{crit}$ above which the system becomes kink stable is to locate the $\tau$ at which the fitted function becomes negative. This process is repeated to find $\tau_{crit}$ as a function of $\tilde{C}_{nm}$.

The procedure can be easily modified to study the kink instability in terms of a single parameter $\alpha$ or $q$ in the cases of Lundquist and GH models. Moreover, the radial perturbation $\xi_r$ that solves the Euler-Lagrange equation (Eq. \ref{eq_euler}) can be plotted to get more insight into the behavior of the instability, and it should be studied more in depth in future research. Any other desired magnetic profile can also be analyzed with this method.

\section{Results and Discussion} \label{s:discussion}

\begin{figure}[h!]
   \centerline{\includegraphics[width=0.8\textwidth]{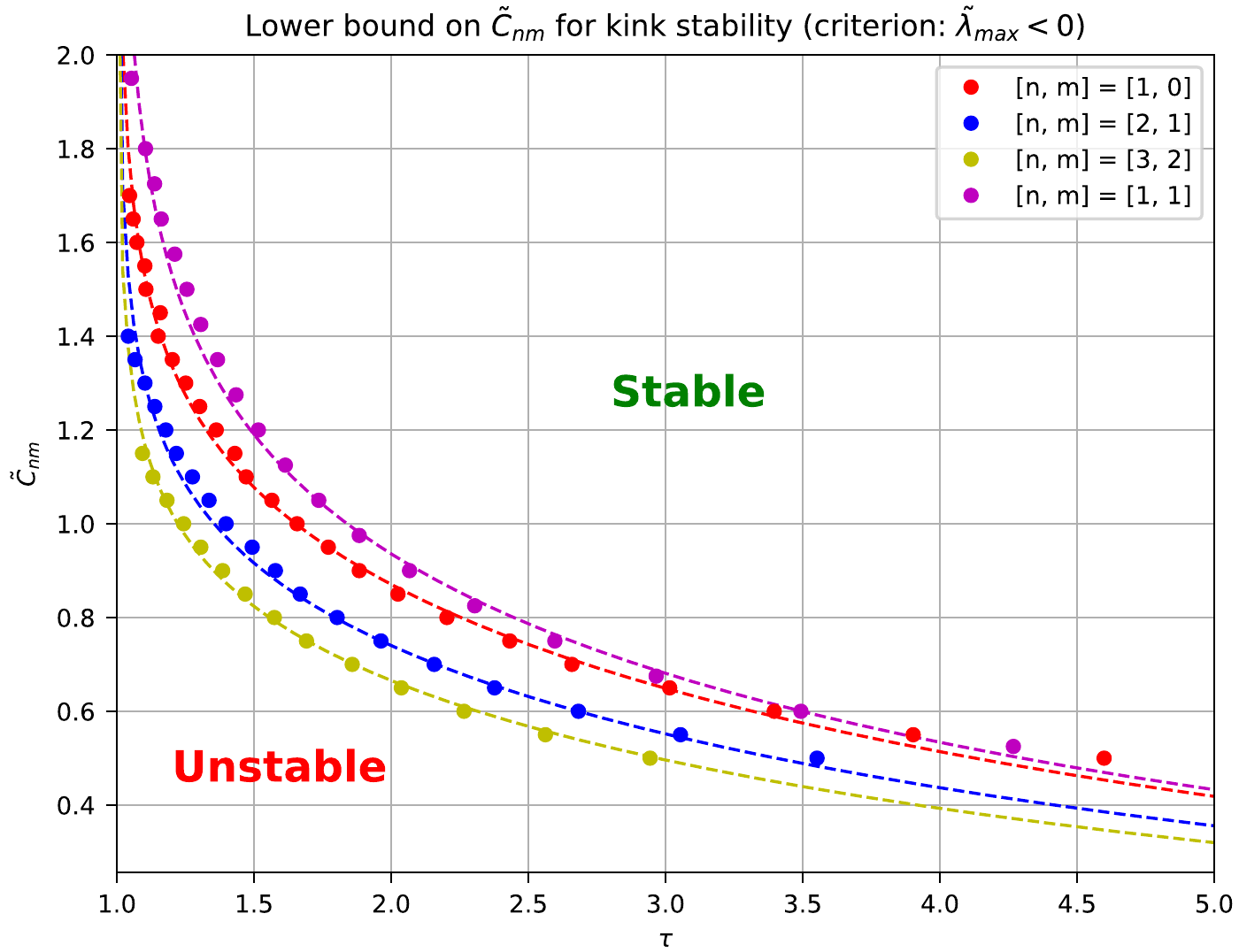}}
   \caption{Plot of the points and the fitted Weibull distributions (dashed lines) symbolizing the boundary values of $\tilde{C}_{nm}$ at fixed $\tau$ between the stable and unstable regimes of the MFR.}
   \label{fig_stab}
\end{figure}

Figure \ref{fig_stab} shows the results of the numerical analysis for the kink instability of the CC model. The points correspond to the minimum $\tilde{C}_{nm}$ above which the system becomes kink stable for each $\tau$, for the different cases $[n, m]$, obtained as explained in Section \ref{s:Numerical analysis}. The linear stability analysis of the Lundquist magnetic profile has resulted in a stability threshold of $\alpha_{crit} = 3.2$, so $\alpha < \alpha_{crit}$ implies that the system is kink stable. Likewise, the uniformly-twisted Gold-Hoyle model is kink stable if $q < q_{crit}$, with $q_{crit} = 1.2$.

It is observed in Figure \ref{fig_stab} that, for the CC model, large values of $\tau$ and $\tilde{C}_{nm}$ are likely to be kink stable since it has already been noted that they make the twist $Q$ decrease. Lundquist and GH models are kink stable for small $\alpha$ or $q$.

A modified Weibull distribution \citep{rinne_weibull_2009} has been fitted to each of the analyzed CC cases, with parameter $\gamma = 1$ in the expression
\begin{equation} \tilde{C}_{nm}(\tau) = \frac{\rho\beta}{\eta} \left( \frac{\tau - \gamma}{\eta} \right)^{\beta-1} e^{-\left(\frac{\tau-\gamma}{\eta}\right)^\beta}, \label{eq:weib} \end{equation}
where the fitted values of $\beta, \eta, \rho$ for each $[n, m]$ are shown in Table \ref{table_stab}. The adjusted functions are plotted in Figure \ref{fig_stab} as dashed lines. They provide a good estimate of the stability limit curve for $\tau \in [1.0, 4.0]$. It is clearly seen that $[1, 1]$ shows the smallest stability range, while the cases of the form $[k+1, k]$ become increasingly stable to kink as $k$ increases. Future exploration of more $[n, m]$ cases could provide an expression relating the fitted parameters with the $n, m$ values considered.

\begin{table}[h!]
    \centering
    {\renewcommand{\arraystretch}{1.1}
    \begin{tabular*}{\textwidth}{cc}
    \textbf{Model} & \textbf{Stability range} \\ \thickhline
    CC $[1, 0]$ & $\tilde{C}_{10} > \tilde{C}_{10,crit}(\tau) =$ (Eq. \ref{eq:weib} with $\gamma = 1, \beta = 0.8400, \eta = 5.7, \rho = 5.6382$) \\
    CC $[2, 1]$ & $\tilde{C}_{21} > \tilde{C}_{21,crit}(\tau) =$ (Eq. \ref{eq:weib} with $\gamma = 1, \beta = 0.8400, \eta = 5.7, \rho = 4.7935$) \\
    CC $[3, 2]$ & $\tilde{C}_{32} > \tilde{C}_{32,crit}(\tau) =$ (Eq. \ref{eq:weib} with $\gamma = 1, \beta = 0.8400, \eta = 5.7, \rho = 4.3101$) \\
    CC $[1, 1]$ & $\tilde{C}_{11} > \tilde{C}_{11,crit}(\tau) =$ (Eq. \ref{eq:weib} with $\gamma = 1, \beta = 0.8032, \eta = 5.8, \rho = 6.1000$) \\
    Lund. & $\alpha < \alpha_{crit} = 3.2$ \\
    GH & $q < q_{crit} = 1.2$ \\
    
    \end{tabular*}}
    \caption{Summary of the kink stability thresholds obtained.}
    \label{table_stab}
\end{table}

Studies of the kink instability usually give the stability threshold in terms of the critical total twist angle of the MFR, $\Phi_c$. This variable is related to the twist $Q$ used here by $\Phi = QL$, where $L$ is the axial length. The equations in Table \ref{table_eqs} show that $\Phi$ is proportional to $\frac{L}{R}$ for all the models under study, and therefore thinner and/or longer MFRs can remain kink stable with higher twist. The comparisons with previous results are not straightforward since the methods and conditions used in each case are different.
    \begin{itemize}
        \item For a uniformly twisted infinite MFR in an incompressible plasma, \cite{dungey_twisted_1954} showed through a normal mode analysis that the critical twist angle is $\Phi_c = 2\frac{L}{R}$, and it can be contrasted with the stability threshold found in this work for the uniformly-twisted GH model: $\Phi_c = q_{crit}\frac{L}{R} = 1.2\frac{L}{R}$, which implies that MFRs become unstable with lower twist. The differences between the two analyses should be explored more in depth to account for possible destabilizing effects.
        \item \cite{hood_critical_1981} found the well-known threshold of $\Phi_c = 2.5\pi$ for line-tied MFRs described by the GH model, which coincides with the experimental result found in \cite{myers_dynamic_2015} in a laboratory set-up specifically modified to resemble solar line-tied MFRs, so the instability does not depend on the aspect ratio $L/R$ when the line-tying condition is assumed.
        \item The constant-$\alpha$ FF Lundquist field has been found to be kink unstable for $\alpha_{crit} = 3.2$. This result is very close to the threshold of \cite{voslamber_stability_1962}, who considered the MFR enclosed by a perfectly conducting rigid wall and obtained $\alpha_{crit} = 3.176$. This result will be discussed further in Section \ref{sec:forces}.
    \end{itemize}

\subsection{Rotations and the Kink Instability} \label{s:rotations}

The helical kink instability makes the axis of a twisted structure become a helix itself. The results of the stability analysis developed in this paper provide an indicator of the range of parameters for which differently modeled MFRs become kink unstable. 

The kink instability is already regarded as a possible explanation for the rotation of $\delta$-spots in the rise of MFRs through the photosphere \citep{kazachenko_sunspot_2010, vemareddy_sunspot_2016, knizhnik_role_2018}. However, this process has not yet been sufficiently explored to account for rotations in the lower-middle corona and heliosphere that have been observed \citep{yurchyshyn_rotation_2009, vourlidas_first_2011}.

The accumulation of poloidal magnetic flux during the first stages of the evolution of a CME could modify the internal twist distribution and physical parameters of the MFR, and this change could drive the onset of kink instabilities that would be seen as rotations in remote sensing coronagraphs \citep{vourlidas_first_2011, nieveschinchilla_remote_2012}. This contrasts with the phenomenon of CME deflection \citep{kay_global_2015} that results from the interaction with the ambient solar wind. On this matter, nonlinear simulations should be done in each case to study how the instabilities evolve in the long run. 

The critical twists found in Table \ref{table_stab} could be used along with observational studies in order to infer which MFRs are susceptible to rotate. For example, the study made by \cite{wang_twists_2016} on 126 MCs of Lepping's list \citep{lepping_summary_2006} with the velocity-modified GH model, showed that interplanetary MFRs have a total twist angle $\Phi$ bounded by $0.2\frac{L}{R} < \Phi < 2\frac{L}{R}$ with an average of $\Phi = 0.6\frac{L}{R}$. The GH threshold of $\Phi_c = q_{crit}\frac{L}{R} = 1.2\frac{L}{R}$ found in the present work denotes that interplanetary MFRs are kink stable on average, but there is a large amount of events with $1.2\frac{L}{R} < \Phi < 2\frac{L}{R}$ that would be kink unstable. Further study of these events and of possible signatures of rotation would allow us to gain better insight into the relation between rotations and the occurrence of the kink instability in the interplanetary medium.

Moreover, analyzing the kink stability of MFR models with multiple free parameters provides physically meaningful constraints on them. It is the case for the CC model: stable MFRs with the CC magnetic field (for $[n, m] \in \{[1, 0], [2, 1],$ $[3, 2], [1, 1]\}$) should satisfy the stability thresholds of Table \ref{table_stab}.

\subsection{Magnetic Forces and the Kink Instability} \label{sec:forces}

The misalignments between $\vec{j}$ and $\vec{B}$ of the MFR models under study were calculated in Section \ref{sec:FF} and expressed in Table \ref{table_eqs}. While the inside of Lundquist and GH MFRs is completely FF, the misalignment for the CC model showed different behaviors depending on the pair $[n, m]$ and the parameters $\tilde{C}_{nm}, \tau$, as shown in Figure \ref{fig_FF}. This raised the question of whether there exists any relation between the magnetic forces that act on an MFR and the kink stability. On the basis of the results yielded by the numerical analysis, the following remarks can be made:
\begin{itemize}
    \item Figure \ref{fig_stab} shows that the CC $[1, 1]$ case has a smaller stability range than the cases $[k+1, k]$, $\forall k$. The main difference between them is that $[1, 1]$ is non-twisted and outward-NFF around the core, while $[k+1, k]$ cases are FF close to the axis. This implies that in the unstable regime of the $[1, 1]$ case, there are outward magnetic forces around the core and inward magnetic forces close to the boundary, suggesting that the presence of magnetic forces in opposite directions within the MFR could have a strong destabilizing effect.
    
    \item Figure \ref{fig_FF} shows that larger $\tilde{C}_{nm}$ makes the boundary of the MFR be more outward-NFF. Similarly, larger $k$ in cases $[k+1, k]$ makes the edge be more outward-NFF (as can be seen in Figure \ref{fig_FF}b, \ref{fig_FF}d and \ref{fig_FF}e). The results of the kink instability analysis in Figure \ref{fig_stab} show that large $\tilde{C}_{nm}$ and large $k$ in $[k+1,k]$ cases make the stability range bigger. This suggests that outward magnetic forces near the edge could have a kink stabilizing effect.
    
    \item Smaller values of $\tau$ make the MFR be more inward-NFF around the boundary (as can be seen in Figure \ref{fig_FF}a, \ref{fig_FF}b and \ref{fig_FF}c), while Figure \ref{fig_stab} shows that the system becomes more kink unstable.
    
    \item Among the cases $[k+1, k]$ that have been analyzed, $[1, 0]$ has non-zero twist at the core. The twist vanishes at $r = 0$ for the rest of them (as shown in Figure \ref{fig_Q}). On the other hand, Figure \ref{fig_stab} shows that $[1, 0]$ is less stable than $[k+1, k]$, for $k \geq 1$. This suggests that the presence of twist close to the axis could have a kink destabilizing effect.

    \item Some studies give arguments in favor of the stability of constant-$\alpha$ FF fields (i.e. with $\nabla \times \vec{B} = \alpha_0 \vec{B}$, where $\alpha_0$ is constant). For example, nonlinear simulations of kink unstable magnetic flux tubes in solar active regions, starting with a NFF magnetic equilibrium, have been shown to evolve into constant-$\alpha$ FF solutions \citep{linton_nonlinear_1998}. In fact, if the magnetic energy of a system is a minimum for a given total magnetic helicity, then $\nabla \times \vec{B} = \alpha_0 \vec{B}$ for some constant $\alpha_0$ \citep{woltjer_theorem_1958}. However, it has been found in this work that Lundquist MFRs, which are constant-$\alpha$ FF, become kink unstable for $\alpha > \alpha_{crit} = 3.2$. This was further addressed in \cite{voslamber_stability_1962} where $\alpha_{crit}$ was shown to be 3.176 for a Lundquist MFR enclosed by a perfectly conducting rigid wall.

    \item In the stability limit, both the FF Lundquist model and the NLFF GH model have a larger average twist than the CC cases, for almost all $\tau$. This implies a possible relation between the force-freeness of a structure and an enhanced stability to kink. 
\end{itemize}

In the future, the relation between the occurrence of the kink instability and the presence of inward or outward magnetic forces around the core or near the edge should be studied. The kink stability of other NFF MFR models should be analyzed to see if these results are corroborated. In addition, further theoretical research is required on the different sources of non-force-freeness and the nature of the magnetic forces that they would induce on MFRs (inward or outward, around the core or close to the edge), as well as on the effects they would have on the onset of the kink instability. 

\subsection{Reversed Chirality Scenario and the Kink Instability}

Another phenomenon that has been observed to occur for certain values of the CC and Lundquist parameters is the change of the sign of $B_y$ at some distance from the MFR axis. In general, assuming that the MFR magnetic field components around the axis are positive (with left-handed (L) chirality), three scenarios can occur in which the sign of the magnetic field components changes some distance away from the core. The three possibilities are described in Table \ref{tab:reversed}.

\begin{table}[h!]
    \centering
    {\renewcommand{\arraystretch}{2.0}
    \begin{tabularx}{\textwidth}{c|cccc|c}
        \multirow{2}{*}{\textbf{Option}} & \multicolumn{4}{c|}{$\vec{B}$ \textbf{properties}} & \multirow{2}{*}{\textbf{Sketch}} \\ \cline{2-5}
        & \textbf{Part} & $\mathbf{B_y}$ & $\mathbf{B_\varphi}$ & \textbf{Chirality} & \\ \hline
        \multirow{2}{*}{(a)} & Int. & + & + & L & \multirow{2}{*}{\includegraphics[width=2.9cm]{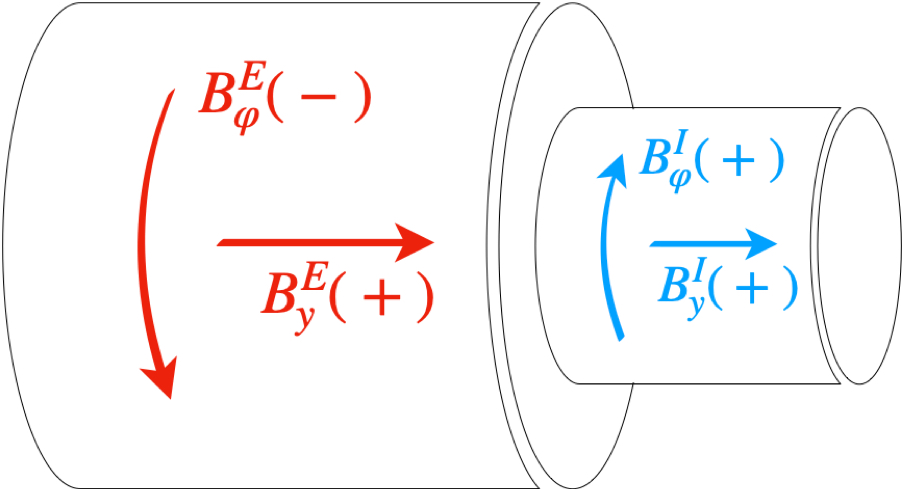} \includegraphics[width=2.9cm]{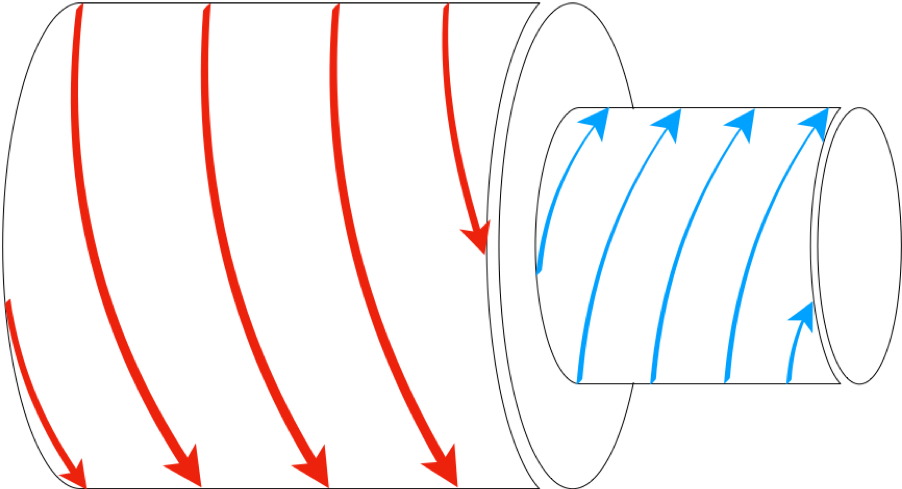}} \\ \cline{2-5}
        & Ext. & + & - & R & \\ \hline
        \multirow{2}{*}{(b)} & Int. & + & + & L & \multirow{2}{*}{\includegraphics[width=2.9cm]{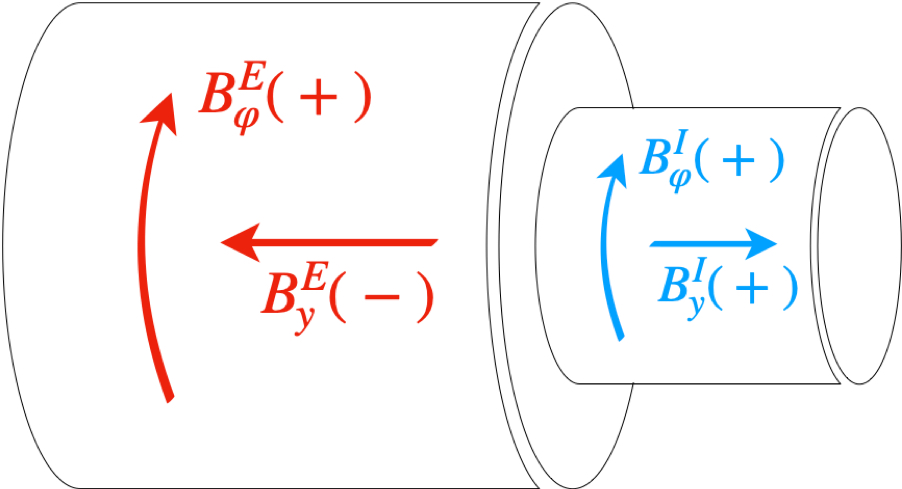}\includegraphics[width=2.9cm]{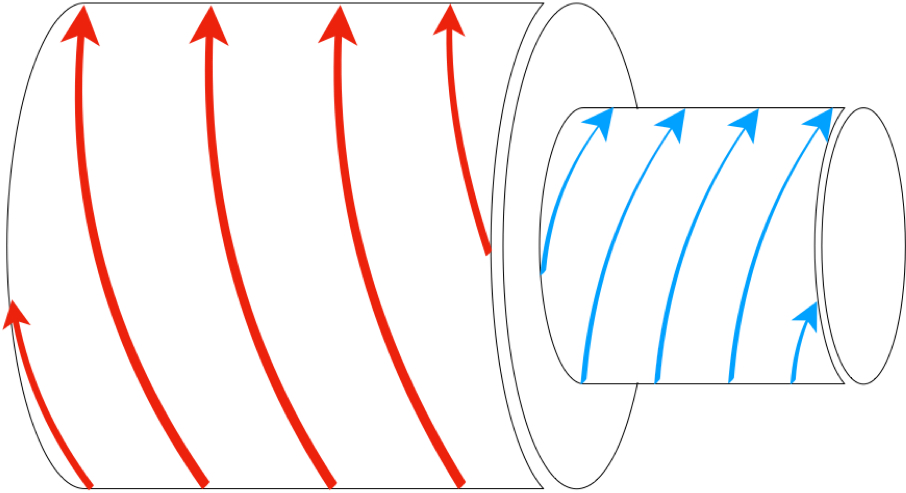}} \\ \cline{2-5}
        & Ext. & - & + & R & \\ \hline
        \multirow{2}{*}{(c)} & Int. & + & + & L & \multirow{2}{*}{\includegraphics[width=2.9cm]{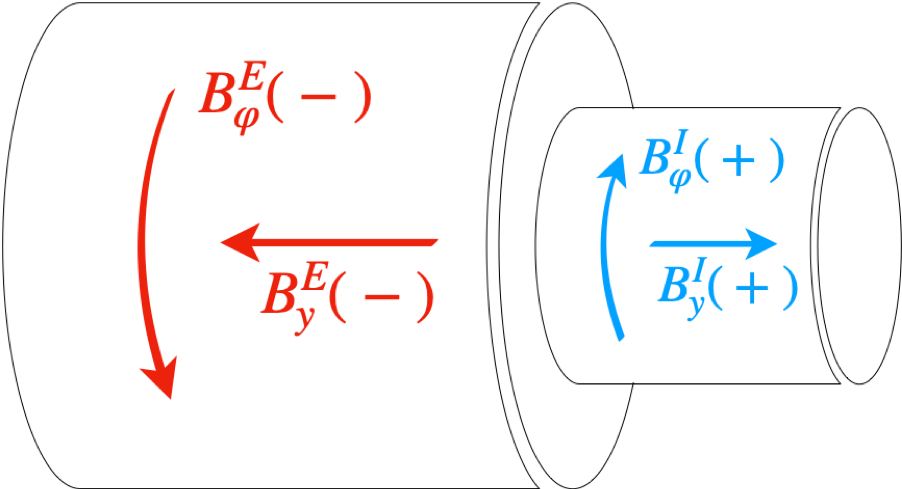}\includegraphics[width=2.9cm]{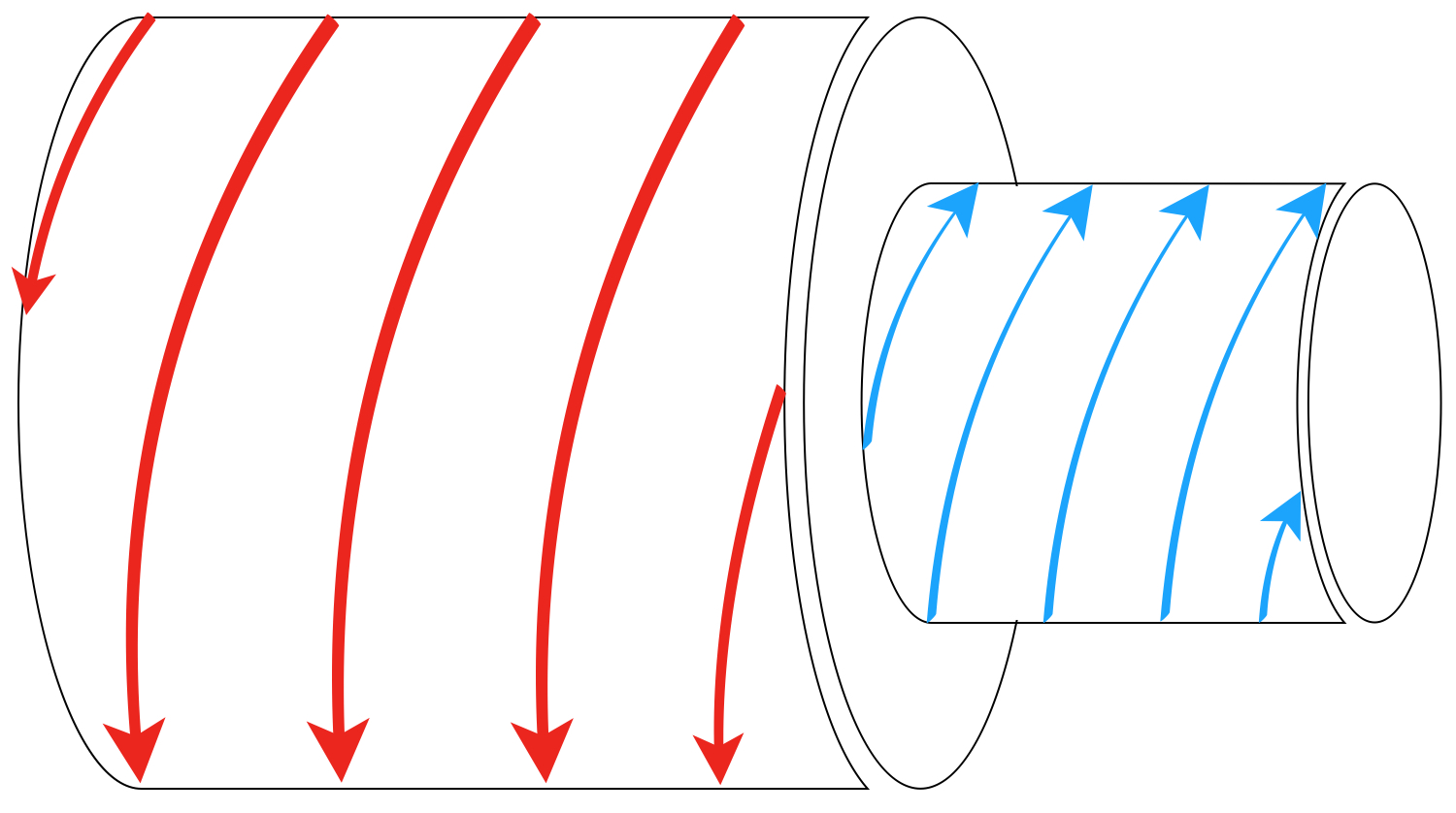}} \\ \cline{2-5}
        & Ext. & - & - & L & \\ 
    \end{tabularx}}
    \caption{Description of the scenarios in which the sign of the magnetic field components changes at some distance from the MFR axis. The internal (int.) axial and poloidal magnetic fields are assumed to be always positive, so that the inner part of the MFR is left-handed (L). The different possibilities of change in the signs of the external (ext.) $B_y$ and $B_\varphi$, and their corresponding chirality, are listed. Two sketches of each scenario are included to clarify the directions of the magnetic field components and the helices that they generate.}
    \label{tab:reversed}
\end{table}

While the GH model does not admit any change in the sign of $B_y$ or $B_\varphi$, option (b) in Table \ref{tab:reversed} is modeled by Lundquist with $\alpha > 2.4$, and by the CC model with $\tau < 1$, for any $[n, m]$, as was noted in Section \ref{s:B}. This scenario predicts a change of the chirality of the MFR at some distance from the axis, where the twist becomes infinite. In fact, Lundquist also models an additional change in the MFR chirality if $\alpha > 3.83$, since $B_\varphi$ becomes negative. 

The results of the stability analysis have shown that a CC MFR is always kink unstable for $\tau \leq 1$ (see Figure \ref{fig_stab}), or in other words, option (b) is always kink unstable for the CC $[n,m]$ pairs that have been studied. This is not the case for the Lundquist model, which becomes unstable only when $\alpha \geq 3.2$, so option (b) can happen while still being kink stable. However, the Lundquist double chirality reversal scenario ($\alpha > 3.83$) is kink unstable.

The fact that the Lundquist model remains kink stable in the reversed chirality scenario raises the question of how magnetic flux can be added in opposite directions during the first stages of the CME evolution. Some events have already been observed that support this hypothesis \citep{cho_comparison_2013, vemareddy_successive_2017}. Actually, a similar magnetic configuration, the reversed field pinch (RFP), which is an axisymmetric toroidal structure with a change of the $B_y$ sign (option (b) in Table \ref{tab:reversed}), has been subject of extensive research for the confinement of laboratory plasmas, due to its being a minimum-energy state of the system and stable against localized MHD instabilities \citep{schwarzschild_reversed-field_1981}. 

Some theoretical studies have been done on the physical consequences of the options of Table \ref{tab:reversed}. For example, \cite{einaudi_ideal_1990} found important differences between magnetic fields with and without an inversion in the $B_y$ sign, since the inversion implied a much higher critical twist for stability, as well as a different nature of the instabilities that would involve more drastic consequences for the non-linear evolution. 

Nevertheless, further research on the stability and origin of these MFR scenarios is needed to understand the physical processes that would be occurring and how they could have been generated. Although the models that have been studied in the present work correspond only to option (b), the consideration of an additional term of the axial current density of the CC model (Eq. \ref{eq:current}) would allow an inversion of the $B_\varphi$ sign, and thus options (a) and (c) could be explored in the future with the numerical method that has been developed here. 

\subsection{Expansion and the Kink Instability}

\begin{figure}[h!]
   \centerline{\includegraphics[width=0.9\textwidth]{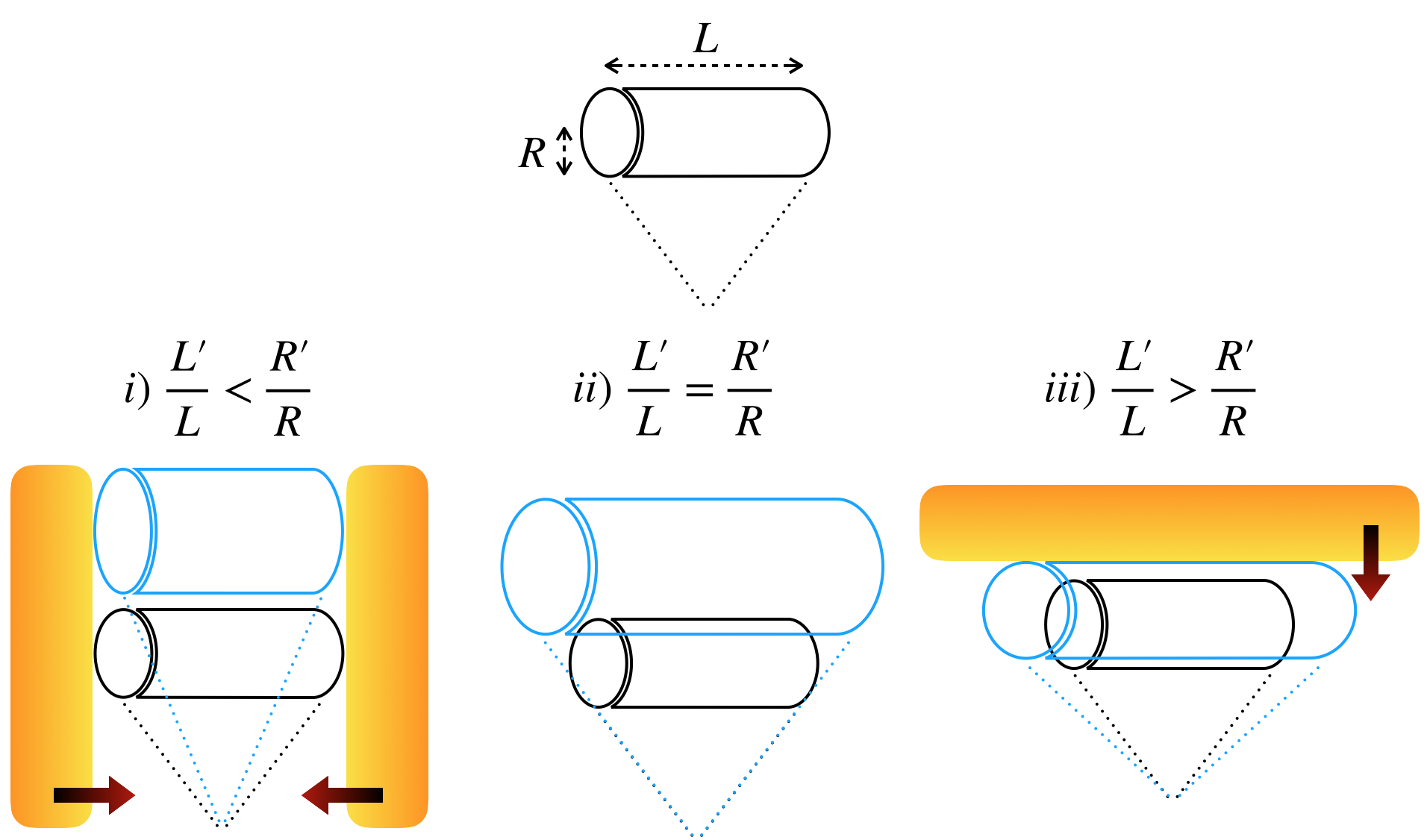}}
   \caption{Top: schematic representation of an MFR, with the radius $R$ and the axial length $L$ indicated. Bottom: examples of different types of interaction that could be causing each of the three aforementioned expansion scenarios. The black MFR is the original one, while the light blue MFR represents the structure after expansion with radius $R'$ and axial length $L'$. The dotted auxiliary lines connect the MFR ends to its point of origin at the Sun. The orange rectangles are abstractions of possible interacting agents in the heliosphere, moving towards the MFR in the directions indicated by the arrows.}
   \label{fig_exp}
\end{figure}

Regarding the evolution of interplanetary MFRs, some conclusions can also be inferred from the results of the stability analysis of the CC model. In Section \ref{s:expansion}, Eq. \ref{eq:expansion} shows the relation between the parameters $B_y^0, \tau, \tilde{C}_{nm}$ of an MFR with radius $R$ and axial length $L$, and the value of the parameters ${B_y^0}', \tau', \tilde{C}_{nm}'$ when it evolves into an MFR with radius $R' > R$ and axial length $L'$. Three possible scenarios can be identified by taking into account the results in Figure \ref{fig_stab} and the fact that, upon expansion, a CC MFR has constant $\tau$, so that any change in the kink stability will be given exclusively by the variation of $\tilde{C}_{nm}$ during its interplanetary journey ($\tilde{C}_{nm}' = \frac{L'}{L}\frac{R}{R'}\tilde{C}_{nm}$):
\begin{enumerate}
    \item $L'/L < R'/R$: the rate of expansion of the axis length is smaller than the radial growth rate. In this scenario, $\tilde{C}_{nm}' < \tilde{C}_{nm}$, so an initially stable MFR can become unstable at some point of its propagation. 
    \item $L'/L = R'/R$: this scenario corresponds to that of self-similar expansion. $\tilde{C}_{nm}' = \tilde{C}_{nm}$, so no change in the kink stability is produced.
    \item $L'/L > R'/R$: the rate of expansion of the axis length is bigger than the radial growth rate. In this case, $\tilde{C}_{nm}' > \tilde{C}_{nm}$, so even an initially unstable MFR could become kink stable in the course of its evolution.
\end{enumerate}
These results suggest that different types of interaction may change the stability of a CME and cause its rotation or its stabilization. Examples of each of the three scenarios are sketched in Figure \ref{fig_exp}. In case (i), an interaction taking place along the sides of a CME could slow down the axial length growth. This would have a destabilizing effect on it and could cause its rotation, since $\tilde{C}_{nm}' < \tilde{C}_{nm}$. In case (ii), without any interaction, the MFR would expand self-similarly and its kink stability would not be affected because $\tilde{C}_{nm}' = \tilde{C}_{nm}$. In case (iii), an interaction acting on the front of the CME could slow down the radial growth. This would have a stabilizing effect, since $\tilde{C}_{nm}' > \tilde{C}_{nm}$.

The next step should be to test these hypotheses with observational data. This could be done by inferring the ratios $L'/L$ and $R'/R$ from the parameters that have been used for the characterization of expanding CMEs until now, e.g. the angular width of the external envelope of the CME from the lateral and from the axial perspectives, i.e. $AW_L$ and $AW_D$, respectively \citep{cabello_first_2016, cremades_asymmetric_2020}, or the lateral and radial expansion speeds, $V_{lat}$ and $V_{front}$ \citep{balmaceda_expansion_2020}. Then, single events with $L'/L > R'/R$ (scenario (iii)) could be studied to see if they are likely to display rotation signatures during their evolution, for example.  

Future research should include the exploration of the relation between the expansion and the kink stability of MFRs described by other models. Further studies on the different types of interaction that could take place and their effect on the onset of CME rotations, as well as more comparisons with observational studies, are still needed to improve the current understanding of the evolution of CMEs and their kink stability.

\subsection{Further Remarks}

Figure \ref{fig_stab} clearly shows that bigger values of $k$ for the CC cases of the form $[k+1, k]$ imply a larger range of stable parameters. On the other hand, in Section \ref{sec:twist}, it was observed that larger $k$ implied a smaller growth rate of the twist around the core, showing a smaller average twist and adopting more of a stage-like distribution (the term is used here to describe distributions that show different behavior of the twist around the core and in the outer shell, that is, almost uniform around the axis and abruptly increasing close to the edge). This suggests that an MFR with a stage-like twist distribution is more stable than others with the same $B_y$ and $B_\varphi$ at the boundary. In the case of the eruption of a CME that requires the triggering of the kink instability, this could suggest that an MFR with continuously distributed twist is a more likely initiation scenario than the corresponding more stable stage-like distribution (associated to preexisting MFR models, e.g. \citealp{kopp_magnetic_1976, titov_basic_1999, longcope_quantitative_2007}). However, the twist distribution of MFRs generated by different initiation processes (e.g. the breakout vs. flux cancellation models) needs further exploration, as well as the kink stability of MFR models with more diverse twist profiles.

Finally, the study of the kink instability of MFRs can also help to review the choice of the parameters that is usually made for some models, as well as to find physically meaningful constraints for models that include multiple free parameters, in order to avoid the occurrence of the instability, as explained in Section \ref{s:rotations}. For example, this analysis has shown that $\alpha$ in the Lundquist model can be varied around $2.4$ before becoming unstable for $\alpha > 3.2$ (as discussed in \citealp{demoulin_re-analysis_2019}). Additionally, it has been found that the typical value $\tau = 1$ that is often used to fit events with the CC model results in kink instability, for the considered $[n, m]$ pairs.

\section{Summary and Conclusions} \label{s:Summary}

This article analyzes the linear kink stability in MFRs modeled by the CC, Lundquist and GH models, following the procedure in \cite{linton_helical_1996}. A numerical method has been developed that is able to find the linear stability range of parameters for any given magnetic configuration. The properties of interest of each of the studied models are described. The results of the numerical procedure are summarized in Table \ref{table_stab} and Figure \ref{fig_stab}. Some of the main conclusions that have been drawn from this analysis are: 
\begin{itemize}
    \item The kink instability could be the cause of the rotations of CMEs that are produced due to their internal magnetic configuration. The results of the analysis performed provide an indicator of the range of parameters for which differently modeled MFRs become kink unstable, being thus susceptible to start rotating.
    
    \item The study of the kink stability of MFRs subject to different magnetic forces suggests that they have a relevant relation to the onset of the instability. In particular, the presence of magnetic forces in opposite directions within the MFR appears to have a strong destabilizing effect, while outward magnetic forces near the boundary seem to be connected to more stable structures.
    
    \item Lundquist MFRs become unstable when the parameter exceeds a certain threshold, showing that constant-$\alpha$ FF fields are also subject to instabilities.
    
    \item The reversed chirality scenario has turned out to be stable for Lundquist MFRs, and kink unstable for the studied CC $[n, m]$ pairs.
    
    \item The evolution of the kink instability of expanding CMEs modeled by the CC model depends on the relation between $L'/L$ and $R'/R$: $L'/L < R'/R$ implies that an initially stable MFR can become unstable at some point of its propagation, $L'/L = R'/R$ indicates self-similar expansion and no change in the kink stability, and $L'/L > R'/R$ makes the structure become more kink stable.
\end{itemize}

Avenues for further research concerning the kink stability in relation to MFR rotations, magnetic forces, the reversed chirality scenario and the expansion, among others, are suggested in Section \ref{s:discussion}. 

This article provides theoretical background to address questions about the impact of the internal twist distribution of MFRs on the kink stability and its relation with the internal magnetic forces distribution in a dynamically expanding structure. In this regard, whether or not all solar-heliospheric flux ropes are alike is still an open question that can be related not only to their formation but also to evolutionary processes. New missions like Parker Solar Probe or Solar Orbiter will provide observations of unexplored areas where the pristine MFRs are less affected by various phenomena that can occur throughout their evolution. Thus, for instance, the reversed chirality scenarios and different boundary conditions for the kink stability analysis, could be studied with the remote-sensing observations in conjunction with the in situ measurements from these missions in the lower-middle corona.

%
\begin{acks}
M. Florido-Llinas thanks CFIS (UPC) and her family for the funding support, and is very grateful to the Heliospheric Physics Laboratory (HSD) at NASA Goddard Space Flight Center for providing the hosting and guidance to carry out this research as part of her bachelor thesis. The work of T. Nieves-Chinchilla is supported by the Solar Orbiter and Parker Solar Probe missions. The work of M.G. Linton is supported by the Office of Naval Research 6.1 program and by the NASA Living With a Star program.
\end{acks}

\begin{conflicts}
The authors declare that they have no conflicts of interest.
\end{conflicts}

%
%

\bibliographystyle{spr-mp-sola}
\bibliography{references_manual.bib}  

%
%
%
%

\end{article} 
\end{document}